\documentclass[aps,prl,reprint,superscriptaddress]{revtex4-2}
\usepackage{graphicx}

\begin{document}


\title{Robust Long Range Magnetic Correlation across Anti-phase Domain Boundaries in Sr$_2$CrReO$_6$}


\author{Bo Yuan}
\affiliation{Department of Physics, University of Toronto, Toronto, Ontario, M5S 1A7, Canada}
\author{Subin Kim}
\affiliation{Department of Physics, University of Toronto, Toronto, Ontario, M5S 1A7, Canada}
\author{Sae Hwan Chun}
\affiliation{Department of Physics, University of Toronto, Toronto, Ontario, M5S 1A7, Canada}
\author{Wentao Jin}
\affiliation{School of Physics, Key Laboratory of Micro-Nano Measurement-Manipulation and Physics (Ministry of Education), Beihang University, Beijing, 100191, China}

\author{C. S. Nelson}
\affiliation{National Synchrotron Light Source II, Brookhaven National Laboratory, Upton, New York, 11973}
\author{Adam J. Hauser}
\affiliation{Department of Physics and Astronomy, The University of Alabama, Tuscaloosa, Alabama, 35487}
\author{F. Y. Yang}
\affiliation{Department of Physics, The Ohio State University, Columbus, Ohio 43210, USA}

\author{Young-June Kim}
\affiliation{Department of Physics, University of Toronto, Toronto, Ontario, M5S 1A7, Canada}

\date{\today}

\begin{abstract}
Anti-site disorder is one of the most important issues that arises in synthesis of double perovskite for spintronic applications. Although it is known that anti-site disorder leads to a proliferation of structural defects, known as the anti-phase boundaries that separate ordered anti-phase domains in the sample, little is known about the magnetic correlation across these anti-phase boundaries on a microscopic level. Motivated by this, we report resonant elastic X-ray scattering study of room temperature magnetic and structural correlation in a thin-film sample of Sr$_2$CrReO$_6$, which has one of the highest $\mathrm{T_C}$ among double perovskites. Structurally, we discovered existence of anti-phase nanodomains of $\sim$15~nm in the sample. Magnetically, the ordered moments are shown to lie perpendicular to the $c$ direction. Most remarkably, we found that the magnetic correlation length far exceeds the size of individual anti-phase nanodomains. Our results therefore provide conclusive proof for existence of robust magnetic correlation across the anti-phase boundaries in Sr$_2$CrReO$_6$.
\end{abstract}


\maketitle

\begin{figure}[tb]
\includegraphics[width=0.5\textwidth]{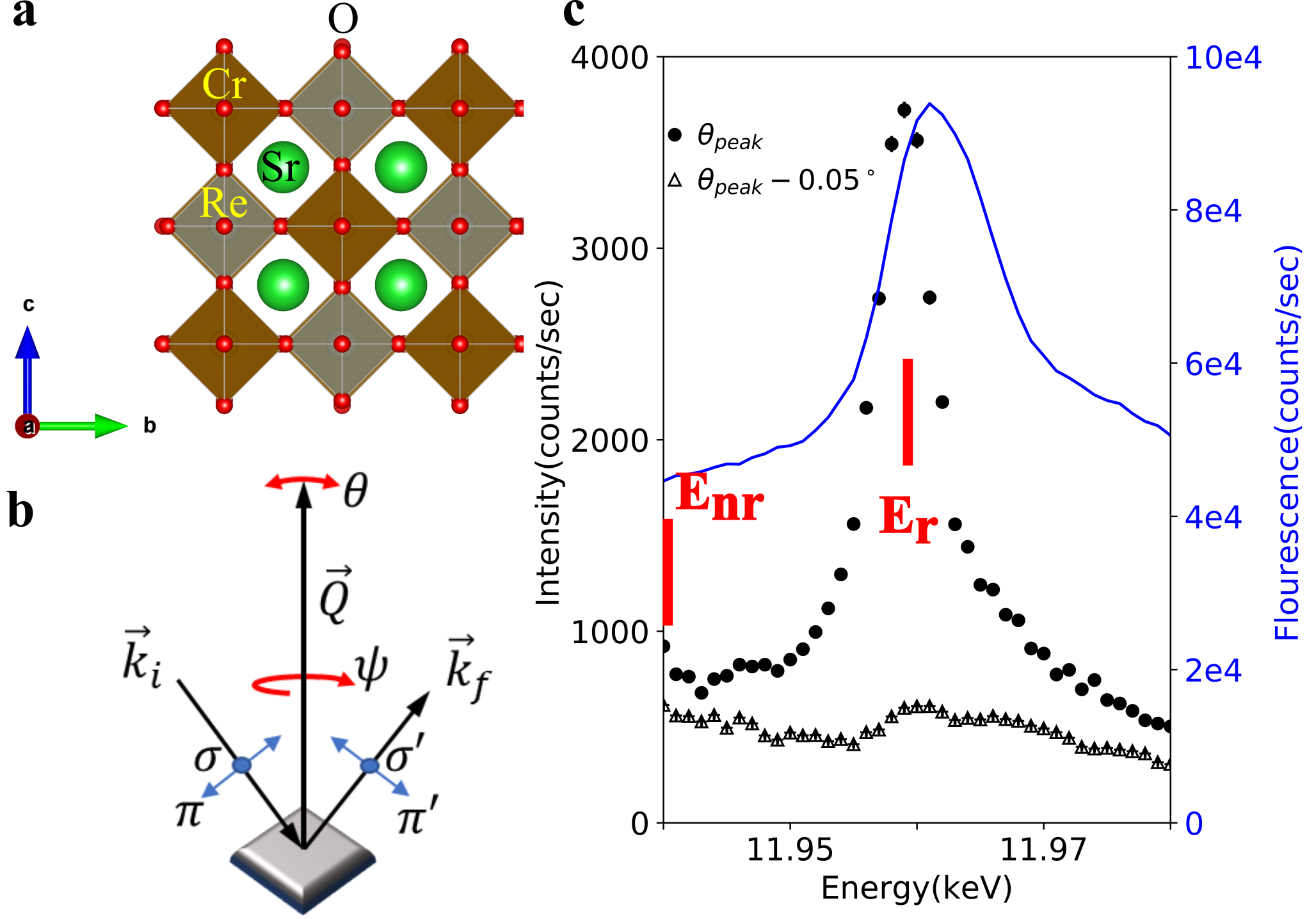}
\caption{(a) Unit cell of an ideal double perovskite. Throughout the paper, we use a cubic notation with $a \approx 7.8$\AA. (b) Experimental setup for resonant elastic X-ray scattering. The angle $\theta$ denotes a rotation within the scattering plane and azimuthal angle $\psi$ denotes a rotation of sample around the momentum transfer $\mathbf{Q}=\mathbf{k}_f-\mathbf{k}_i$. $\psi=0$ is defined as when the reference wave-vector, $\mathbf{Q}_\mathrm{ref}=$(0,0,1) is in the scattering plane. X-ray can be either polarized parallel or perpendicular to the scattering plane, which is denoted by $\sigma$ and $\pi$ polarizations, respectively. Polarizations of out-going X-ray is denoted with a prime. (c) Energy dependence of the $\sigma-\pi^\prime$ channel elastic intensity at $\mathbf{Q}=(1,1,5)$ near Re L2 edge. The energy dependence is obtained for different $\theta$'s. The solid black circles are the energy dependence at the peak $\theta$ value, or $\theta_{peak}$, while open triangles are for the background scan obtained 0.05$^\circ$ away from $\theta_{peak}$.}
\label{fig1}
\end{figure}

Ordered double perovskite (DP), $\mathrm{A_2 B B^\prime O_6}$, has been intensively studied as one of the most promising candidates for spintronic materials. In an ideal DP, the transition metal (TM) B and B$^\prime$ ions are alternately arranged in a rock salt structure, surrounded by corner sharing oxygen octahedra (See Fig.~\ref{fig1}a). A strong hybridization between B and B$^\prime$ ions can lead to a very high magnetic ordering temperature $\mathrm{T_C}$, and a large spin polarization at the Fermi level, giving rise to the large magnetoresistance at room temperature essential for spintronic applications\cite{Serrate_2006}. The most well known example is Sr$_2$FeMoO$_6$ with a $\mathrm{T_C}\sim$420~K\cite{Tomioka2000,Kobayashi1998}, which is later surpassed by Sr$_2$CrReO$_6$\cite{Kato2002, Hauser2012, Asano2004, Teresa2005,GEPRAGS20092001} and Sr$_2$CrOsO$_6$\cite{Krockenberger2007, Samanta2015, Rhazouani2016} with a $\mathrm{T_C}$ of 635~K and 725~K, respectively. However, one important issue that inevitably arises during the synthesis of a DP sample is anti-site (AS) disorder whereby locations of B and B$^\prime$ are exchanged in parts of the sample. One might naïvely expect that the AS disorder occurs randomly as a point defect when a site is occupied by a {\em wrong} atom (B$^\prime$  in B site, for example). While this random occupancy model is useful in distinguishing an ordered DP sample from its fully disordered limit (a solid solution of B and B$^\prime$ in a regular perovskite structure), it is an inaccurate picture for describing the AS disorder in most ordered DP samples. In a more realistic description, AS disorder actually exists as a planar defect known as an anti-phase boundary (APB), which separates two adjacent anti-phase domains (APD) whose B/B$^\prime$ arrangement is reversed. Note that there is no {\em wrong} atom in this picture: Each APD is perfectly ordered and the only {\em defects} are the APB's consisting of B-O-B or B$^\prime$-O-B$^\prime$ bonds. This picture is supported by M\"ossbauer spectroscopy \cite{Greneche2001} as well as transmission electron microscopy (TEM) studies \cite{Navarro2001,Yu2005,Asaka2007}. Recent quantitative scanning TEM study directly imaged the APB in a high-quality Sr$_2$CrReO$_6$ film \cite{Essler2016}. In addition, an X-ray Absorption Fine Structure (XAFS) study \cite{Meneghini2009} also revealed that small APD's persist even in highly disordered DP samples of Sr$_2$FeMoO$_6$.

An important question is the effect AS disorder on a DP's magnetic properties, which directly affect its functionality as a spintronic material. Previous bulk magnetization\cite{Navarro2003,Sanchez2002} and x-ray magnetic circular dichroism (XMCD)\cite{Pal2018} studies revealed that the magnetization is reduced with increasing level of AS disorder. However, these techniques only measure the total magnetic response, which is averaged over all APD's. Thus these studies have not addressed the most important aspect of the question, namely how the neighboring APD's are correlated magnetically.  Since magnetically correlated APD's are much easier to manipulate than random ones, one expects a larger magnetoresistance in the former and hence better performance as a spintronic material\cite{Singh2011, Navarro2001}. The relationship between the APB and magnetic domains have been the topic of a number of studies. In their TEM studies of a Ba$_2$FeMoO$_6$ single crystal sample, Asaka $\it{et\,al.}$ found large APD's of a few hundred nanometers, and a strong pinning of magnetic domain boundaries at the APB\cite{Asaka2007}. However, $\it{quantitative}$ information regarding how magnetic order propagates across APB's on a microscopic level is still lacking.

In this paper, we address this important question using resonant elastic X-ray scattering (REXS), which is a bulk sensitive technique like neutron scattering. However, since its momentum resolution is much better than that of neutron scattering, it is an ideal probe for extracting quantitative information such as the correlation length. We have applied this technique to understand the room temperature magnetic and structural correlations in a thin-film sample of Sr$_2$CrReO$_6$ grown on SrTiO$_3$ with a thickness of 319~nm. We found that our sample consists of small APD's of $\sim$15~nm. Remarkably, the magnetic correlation length is much longer than the average APD size, providing evidence for strong magnetic coupling across the APB's.

In a REXS experiment schematically shown in Fig.~\ref{fig1}b (See Supplemental Materials for further experimental details), magnetic scattering is enhanced by tuning the incident photon energy to an atomic absorption edge. Polarization dependence of the scattered X-ray distinguishes magnetic scattering from the usual charge scattering, thus allowing both the structural and magnetic information to be obtained. For an incident polarization perpendicular to the scattering plane ($\sigma$), charge and magnetic scattering gives rise to out-going X-ray polarizations that are perpendicular ($\sigma^\prime$) and parallel ($\pi^\prime$) to the scattering plane\cite{Lann2008}, respectively. In Fig.~\ref{fig1}c, we show $\sigma-\pi^\prime$ channel resonant intensity at the $\mathbf{Q}=(1,1,5)$ Bragg peak as a function of incident photon energy near Re L2 absorption edge (black solid circle). The x-ray absorption spectrum as a function of energy (blue solid line) is also shown for comparison. The resonance peak energy at $\mathrm{E_r}=11.959~\mathrm{keV}$ occurs slightly below the absorption maximum (so-called white line), consistent with REXS studies on other $5d$ rhenium \cite{McMorrow2003, Hirai2020} and iridium compounds\cite{Kim1329, Liu2011, Biffin2014, Boseggia2013}. However, the elastic intensity does not disappear when the incident energy is tuned to an energy well below the absorption edge (e.g. $\mathrm{E_{nr}}=11.94\mathrm{keV}$). This indicates that in addition to the resonant diffraction intensity, a non-resonant background is also present at this $\mathbf{Q}$. In Fig.~\ref{fig1}c, we also show energy dependence of the intensity slightly off the Bragg peak position, obtained by rotating sample angle $\theta$ (see Fig.~\ref{fig1}b for definition) by 0.05$^\circ$ away from the exact location of the Bragg peak. A very different energy dependence is obtained at this new position: the resonant feature is clearly absent while the non-resonant background remains. This indicates that the background has a much broader momentum dependence than the main resonant feature.

\begin{figure}[tb]
\includegraphics[width=0.5\textwidth]{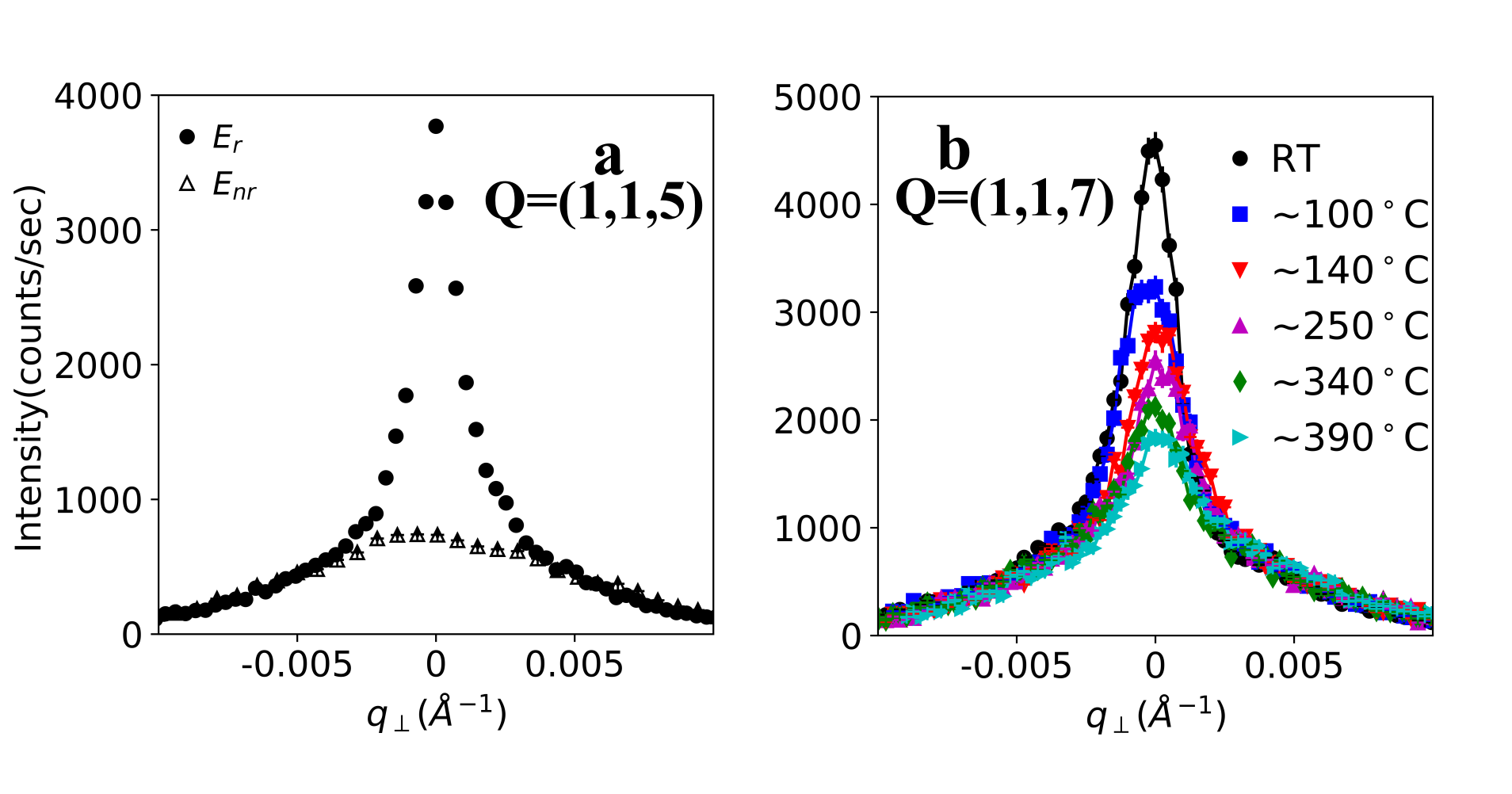}
\caption{(a) Rocking curve ($\theta$) scans at $\mathbf{Q}=(1,1,5)$ in the $\sigma-\pi^\prime$ channel. The $\theta$ values are converted to momentum transfer in $\AA^{-1}$ perpendicular to $(1,1,5)$. $E_r=11.959$~keV and $E_{nr}=11.94$~keV as shown in Fig.~\ref{fig1}c. (b) Temperature dependence of the rocking curves at $\mathbf{Q}=(1,1,7)$ and $E_r$ in the $\sigma-\pi^\prime$ channel. All curves have been shifted horizontally to match the peak positions. Temperature readings were strongly fluctuating during the measurement depending on position of the tip of the thermocouple. Values given in the legend should therefore be taken as a rough estimation of the sample temperature.}
\label{fig2}
\end{figure}

To further illustrate this, we carried out $\theta$-scans using the resonant ($\mathrm{E_r}$) and non-resonant ($\mathrm{E_{nr}}$) incident energies. The resultant rocking curves are shown in Fig.~\ref{fig2}a where the horizontal unit has been converted to momentum transfer perpendicular to $\mathbf{Q}=(1,1,5)$, or $q_\perp$. For the resonant energy $\mathrm{E_r}$ (solid circle), the line shape consists of a very sharp feature on top of a broad background. On the other hand, only a broad background remains when the non-resonant energy $\mathrm{E_{nr}}$ is used. Combining Fig.~\ref{fig1}c and Fig.~\ref{fig2}a, we conclude that the $\sigma-\pi^\prime$ channel elastic intensity near $\mathbf{Q}=(1,1,5)$ consists of two parts: a resonant contribution very sharp in $\mathbf{Q}$ and a non-resonant contribution much broader in $\mathbf{Q}$.

Existence of a sharp resonant peak on top of a diffuse non-resonant background is also observed at other $\mathbf{Q}=$(odd, odd, odd), such as $\mathbf{Q}=(1,1,3)$ (see Supplemental Material) and $\mathbf{Q}=(1,1,7)$. To understand the origins of the resonant and non-resonant contributions to the diffraction intensity, we show rocking curves at $\mathbf{Q}=(1,1,7)$ using the resonant incident energy in Fig.~\ref{fig2}b at different temperatures. As temperature increases towards $\mathrm{T_c}$, the sharp peak is clearly suppressed while the diffuse background is unchanged. Strong temperature dependence and the clear resonance behavior directly indicates the sharp peak's magnetic origin. On the other hand, absence of these characteristics indicates a structural origin of the  non-resonant background. Since Cr and Re magnetic moments are antiparallel in ferrimagnetically ordered Sr$_2$CrReO$_6$, the magnetic unit cell is the same as the structural unit cell shown in Fig.~\ref{fig1}a. Magnetic Bragg peak therefore coincides with the structural Bragg peak. Although charge scattering is ideally only detected in the $\sigma-\sigma^\prime$ channel, some of the charge scattering intensity can leak into the $\sigma-\pi^\prime$ channel as the analyzer is not $100\%$ efficient in removing scattered X-rays with $\sigma^\prime$ polarization. In fact, as we show in Supplemental Materials, $\mathbf{Q}$ scans of diffraction intensity in the $\sigma-\sigma^\prime$ channel and the non-resonant background in the $\sigma-\pi^\prime$ channel show identical lineshapes along all directions, further indicating their common structural origin.

\begin{figure}[tb]
\includegraphics[width=0.5\textwidth]{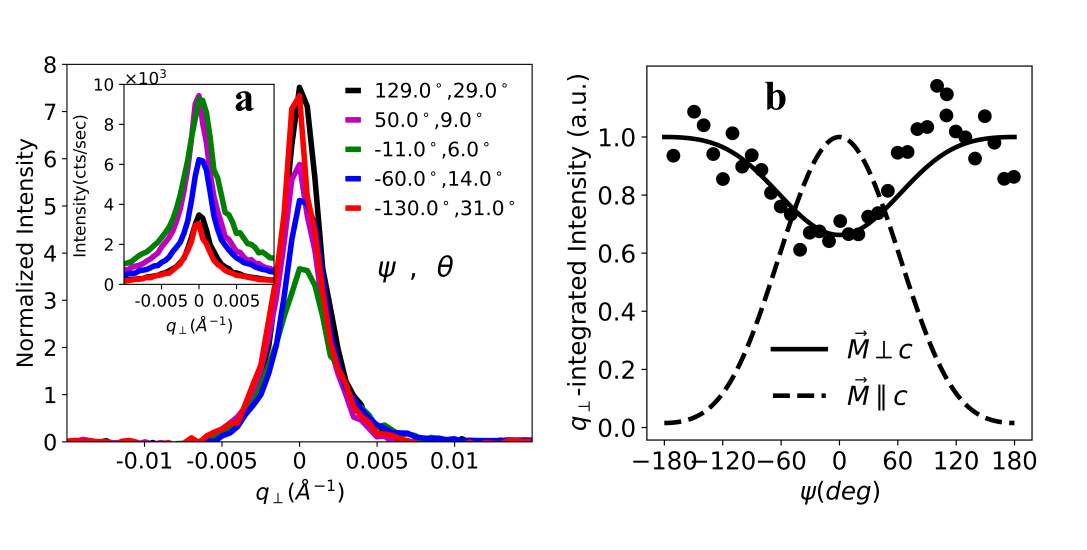}
\caption{(a) $q_\perp$-scans of the $\mathbf{Q}=$(1,1,5) magnetic Bragg peak at different azimuthal angles ($\psi$). The corresponding sample angle ($\theta$) is also indicated. All data are first normalized with respect to the non-resonant diffuse background, which is then subtracted to obtain the magnetic signal. The inset shows the raw $q_\perp$-scans before normalization and background subtraction. (b) Azimuthal dependence of the $q_\perp$-integrated intensity. Solid (dashed) line is the calculated azimuthal dependence at (1,1,5) for ordered moment perpendicular (parallel) to the $c$ axis. The reference wave-vector is $\mathbf{Q}_\mathrm{ref}$=(0,0,1)}
\label{fig3}
\end{figure}

Magnetic scattering intensity in the $\sigma-\pi^\prime$ channel is proportional to the projection of the ordered moment ($\mathbf{M}$) along the direction of scattered wave-vector, or $\left|\mathbf{M}\cdot\mathbf{k}_f\right|^2$\cite{Hill1996}. Therefore, information on direction of $\mathbf{M}$ can be obtained by examining the intensity variation as a function of the azimuthal angle $\psi$ (see Fig.~\ref{fig1}b for definition) as the sample is rotated around $\mathbf{Q}$\cite{Chun2015, Biffin2014, Boseggia2013, Liu2011}. Two problems need to be addressed when extracting magnetic intensity at a given $\psi$. First, the non-resonant structural background needs to be removed to obtain the resonant magnetic contribution. Second, since the X-ray footprint and hence the probed sample volume is enhanced at small incident angle for a thin-film sample, overall intensity is maximized (minimized) when X-ray is incident on the sample from a grazing (normal) direction. This effect is illustrated in the inset of Fig.~\ref{fig3}a: tail of the diffuse structural background is larger at smaller $\theta$. To account for this effect, each $q_\perp$-scan is normalized with respect to the tail at $q_\perp\geq0.005^\circ$ away from the peak position where the resonant magnetic contribution is suppressed. At each azimuthal angle, the non-resonant background was separately measured using a non-resonant incident energy, which is then scaled and subtracted from the normalized $q_\perp$-scans. The resulting $q_\perp$-scans shown in main panel of Fig.~\ref{fig3}a, corresponding to those of the resonant magnetic signal, are then integrated to extract the magnetic intensity as a function of azimuthal angle. In the Supplemental Materials, we present another method to extract the magnetic intensity from energy integration, which gives identical azimuthal dependence shown here.

The resultant azimuthal dependence of magnetic intensity at $\mathbf{Q}=(1,1,5)$ are shown in Fig.~\ref{fig3}b. In a tetragonal crystal such as the Sr$_2$CrReO$_6$ thin-film (see Supplemental Materials), the magnetic moment either points along $c$ or within the $ab$ plane by symmetry. In Fig.~\ref{fig3}b, simulated azimuthal dependence is shown as solid (dashed) lines for an ordered moment pointing perpendicular (parallel) to the $c$ axis. Our data clearly agrees with the solid line, suggesting that the ordered moment is perpendicular to $c$. However, precise ordering direction within the $ab$ plane could not be determined because of averaging over two orthogonal magnetic domains that are present due to tetragonal crystal symmetry. Although an in-plane ordered moment in thin-film Sr$_2$CrReO$_6$ on SrTiO$_3$ has been previously inferred from magnetization measurement \cite{Lucy2014}, our current study demonstrates this directly. Lastly, although we have only shown results obtained near Re L2 edge in the main text, identical results were obtained from our Re L3 REXS experiments (See Supplemental Materials).

\begin{figure}[tb]
\includegraphics[width=0.5\textwidth]{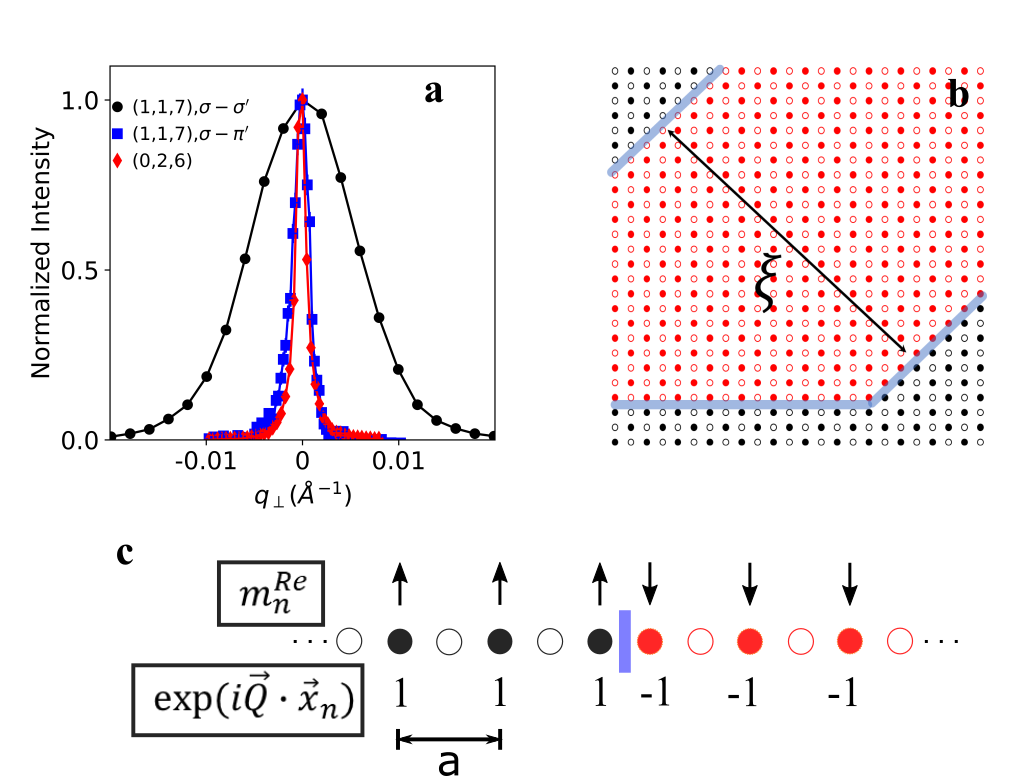}
\caption{(a) Comparison between rocking curves at $\mathbf{Q}=(0,2,6)$ and $\mathbf{Q}=(1,1,7)$. For $\mathbf{Q}=(1,1,7)$, data from both the $\sigma-\pi^\prime$ and $\sigma-\sigma^\prime$ polarization channels are shown. The $\sigma-\pi^\prime$ data at $\mathbf{Q}=(1,1,7)$ is the magnetic intensity obtained by subtracting non-resonant diffuse background. Peak intensity of all rocking curves have been scaled to 1 for comparison. (b) Schematic drawing of APD's separated by APB's (thick lines) in our sample. Open and filled circles denote Cr and Re atoms, respectively. The average domain size is indicated by $\xi$. Red and black colors denote the two types of domains where positions of Cr and Re ions are exchanged. (c) 1D model of the APB (thick vertical line in the middle). The two domains are colored in black and red as in (b). Re magnetic moments, or $m_n^{Re}$ are shown by black vertical arrows above each Re atoms. The phase factor $\exp(i\vec{Q}\cdot\vec{x}_n)$ used to calculate the magnetic Bragg peak intensity at odd $\vec{Q}$'s is shown below each Re atom.}
\label{fig4}
\end{figure}

The most surprising result in our study is the coexistence of a sharp magnetic Bragg peak and a diffuse structural peak at $\mathbf{Q}=$(odd, odd, odd). This important result is emphasized in Fig.~\ref{fig4}a, where we directly compare the rocking curves for the structural and magnetic Bragg peak at $\mathbf{Q}=(1,1,7)$, whose intensities are maximized in the $\sigma-\sigma^\prime$ and $\sigma-\pi^\prime$ channels, respectively. By fitting the structural Bragg peak at $\mathbf{Q}=(1,1,7)$ to a Gaussian, its half width at half maximum (HWHM) is found to be $\kappa=0.0065(6)\AA^{-1}$. In contrast, the structural Bragg peak at $\mathbf{Q}=$(even, even, even) is resolution limited as shown in Fig.~\ref{fig4}a. This is entirely consistent with presence of AS disorder in the sample. To see this, we note that contribution to the scattering amplitude at even and odd $\mathbf{Q}$'s by a nearest neighbour pair of ions, Re, Cr, is proportional to $f_\mathrm{Re}+f_\mathrm{Cr}$ and $f_\mathrm{Re}-f_\mathrm{Cr}$ respectively, with $f_\mathrm{Re},f_\mathrm{Cr}$ being the atomic form factors. Therefore, only odd $\mathbf{Q}$'s will be sensitive to existence of APD in the sample where locations of Re and Cr are exchanged for neighbouring domains (See Fig.~\ref{fig4}b for a schematic drawing of an AS disordered DP crystal with APD's). This leads to selective broadening of structural Bragg peaks at odd $\mathbf{Q}$'s that has been observed in other DP's with APD's\cite{Meneghini2009,Chakraverty2011}. Correlation length extracted from the inverse of broadening in $\mathbf{Q}$ scans therefore gives the average APD size, $\xi$. Using $\kappa=0.0065(5)\AA^{-1}$, this is estimated to be $\xi=\frac{1}{\kappa}=15(2)$~nm in our Sr$_2$CrReO$_6$ sample. By comparing the 319~nm film used here to another 90~nm film (See Supplemental Materials), we confirmed the existence of larger APD's in a sample with lower level of AS disorder.

Having established the presence of APD in Sr$_2$CrReO$_6$, we now move on to discuss its effect on the magnetic order. Naively, one expects proliferation of random structural defects such as the APB's to adversely affect the magnetic order and limit magnetic correlation to $within$ individual APD's. However, this is directly contradicted by our data showing a much sharper magnetic Bragg peak at the $same$ $\mathbf{Q}$'s where the structural Bragg peak is broadened by the finite size of APD's. Remarkably, as shown in Fig.~\ref{fig4}a, the magnetic Bragg peaks have almost the same width as the sharp resolution limited structural Bragg peaks at even $\mathbf{Q}$'s. We therefore arrive at the important conclusion that the magnetic correlation extends far beyond the APD's in Sr$_2$CrReO$_6$.

Microscopically, this implies that two neighbouring APD's to be strongly magnetically coupled in Sr$_2$CrReO$_6$. Since the APB separating the two neighbouring domains consists of 180$^\circ$ Re-O-Re or Cr-O-Cr bonds, simple application of Goodenough-Kanomori rules suggests that two neighbouring domains are antiferromagnetically coupled. Schematically, we illustrate the resulting magnetic and structural arrangement of two neighbouring domains using a heuristic 1D chain model in Fig.~\ref{fig4}c, where solid and open circles denote Re and Cr atoms, respectively (same argument holds true in 3D). Within each domain, the Re and Cr moments are antiferromagnetically coupled, leading to a ferromagnetic arrangement of Re moments shown by black vertical arrows (Cr moment is not shown). Across the APB (denoted by vertical line), all Re moments are flipped with respect to the first domain. The magnetic Bragg peak intensity at a given $\mathbf{Q}$ is proportional to $\sum_n m_n^{Re}\exp(i\mathbf{Q}\cdot\mathbf{x}_n)$, where $\sum_n$ is the sum over all Re-sites and $m_n^{Re}$ is the Re magnetic moment at site $n$. In the 1D model, phase factors $\exp(i\mathbf{Q}\cdot\mathbf{x}_n)$ for odd $\mathbf{Q}$'s (equivalent to $\mathbf{Q}=$(odd, odd, odd) in a 3D model) are shown for the two neighbouring domains in Fig.~\ref{fig4}c. Very importantly, as the phase factor changes sign at the APB, the magnetic moment $m_n^{Re}$ changes sign as well. Magnetic scattering amplitude for two antiferromagnetically coupled APD's therefore add constructively at odd $\mathbf{Q}$'s. This provides a microscopic explanation for why magnetic Bragg peaks at odd $\mathbf{Q}$'s remain sharp in our data while the structural Bragg peaks are broadened.

In conclusion, we have carried out REXS to study room temperature structural and magnetic correlation in a thin-film sample of a high $\mathrm{T_C}$ DP, Sr$_2$CrReO$_6$. We showed directly that the ordered moments lie perpendicular to $c$, which was only inferred from previous magnetization measurements. More importantly, we found a very different structural and magnetic correlation length in the sample, the former is limited by the sizes of the anti-phase domains, or APD's of about 15~nm while the latter far exceeds the average domain size. Our results are consistent with antiferromagnetically coupled APD's. Existence of magnetically correlated APD's has been argued to enhance magnetoresistance of a DP sample, and hence its performance as a spintronic material, due to cooperative spin rotation between different domains\cite{Singh2011}. Our work can be readily extended to systematically study the magnetic and structural correlations in thin-film samples with different levels of AS disorder\cite{Chakraverty2010}. Furthermore, excellent momentum resolution combined with large penetration depth of X-rays also enables one to study the relationship between APD's and magnetic correlations in other DP single crystals.

\begin{acknowledgments}
Work at the University of Toronto was supported by the Natural Science and Engineering Research Council (NSERC) of Canada, Canada Foundation for Innovation (CFI), and Ontario Research Fund (ORF) - Large Infrastructure. Work at the Ohio State University was supported by the Center for Emergent Materials, an NSF-funded MRSEC, under Grant No.
DMR-2011876. This research used beamline 4-ID of the National Synchrotron Light Source II, a U.S. Department of Energy (DOE) Office of Science User Facility operated for the DOE Office of Science by Brookhaven National Laboratory under Contract No. DE-SC0012704.
\end{acknowledgments}

\end{document}